\documentclass[11pt]{article}
\usepackage{amsfonts}
\usepackage{mathrsfs}
\usepackage{graphicx}
\usepackage{amsmath}
\usepackage{amssymb}
\usepackage{mathptmx}
\usepackage{subfigure}
\usepackage{epsfig}
\usepackage{graphicx}
\usepackage{color}
\usepackage{newtxtext,newtxmath}
\usepackage{indentfirst}
\usepackage{hyperref}
\usepackage{setspace}
\usepackage{braket}
\usepackage{slashed}
\usepackage{cite}
\usepackage{multirow}
\usepackage{cancel}
\usepackage{caption}
\usepackage{array}
\usepackage{appendix}
\usepackage{tabularx}
\usepackage{tabulary}
\usepackage{titlesec}
\titleformat{\section}{\normalfont\Large\bfseries\scshape}{\thesection}{1em}{}
\titleformat{\subsection}{\normalfont\large\bfseries\scshape}{\thesubsection}{1em}{}
\usepackage{ulem}
\renewcommand{\arraystretch}{1.2}
\hypersetup{colorlinks=true, citecolor=red, urlcolor=blue, linkcolor=blue}
\begin{document}

\title{Strange pentaquark molecules in QCD sum rules}
\author{Di Ben$ ^{1}\footnote{Contact author: bd@mail.tsinghua.edu.cn}$ and Sheng-Qi Zhang$^{2}\footnote{Contact author: shqzhang@pku.edu.cn}$\\
\\
\normalsize{$^{1}$\it Department of Physics and Center for High Energy Physics, Tsinghua University, Beijing 100084, China}\\
\normalsize{$^{2}$\it Center for High Energy Physics, Peking University, Beijing 100871, China}\\
\\
}

\date{}
\maketitle

\begin{abstract}
We study hidden- and open-strange pentaquark configurations in the hadronic molecular picture with QCD sum rules. Starting from the baryon octet combined with an $s\bar{s}$ pair, we construct 21 interpolating currents with $J=1/2$ and $3/2$. In the analysis, contributions up to dimension 11 are included in the operator product expansion, and both parity-projected spectra and the chiral-even and chiral-odd Lorentz structures are examined. No bound-state solution is found in the $J^P=1/2^-$ sector, although several channels generate resonance masses near nearby thresholds or poorly established strange baryons. In the $J^P=3/2^-$ sector, the $\Sigma\phi$ and $\Xi K^\ast$ currents both support a near-threshold bound-state solution around $2.2$ GeV, suggesting a possible molecular configuration with channel mixing. In contrast, positive-parity sectors are more difficult to identify the corresponding states.
In addition, our results favor the spin-parity assignments $J^P=3/2^-$ for $\Sigma(2250)$, $J^P=1/2^-$ for $\Sigma(2455)$, $\Sigma(2620)$, and $\Xi(2370)$, and $J^P=3/2^+$ for $\Xi(2250)$.
These results provide a systematic QCD sum rule study of strange pentaquark molecular candidates in the strange sector.
\end{abstract}

\section{Introduction}\label{Sec:intro}

The study of nucleon structures moving beyond the traditional quark model, including hadronic molecular, compact pentaquarks, and hybrid states, has emerged as a frontier research area in hadron physics.
The discovery of $J/\psi p$ resonances in 2015 by the LHCb Collaboration presented striking evidence, named $P_c^+(4380)$ and $P_c^+(4450)$, in $\Lambda^0_b\to K^- J/\psi p$ decays~\cite{LHCb:2015yax}.
Further information was reported in 2019; the LHCb Collaboration declared the $P_c^+(4312)$ state and a two-peak structure of the $P_c^+(4450)$ state, which is resolved into $P_c^+(4440)$ and $P_c^+(4457)$~\cite{LHCb:2019kea}.

Within the hadronic molecular picture, these $P_c$ states are well interpreted into $\bar{D} \Sigma_c^\ast$ and $\bar{D}^\ast \Sigma_c$ bound states.
Similarly, systems containing a hidden-charm quark pair can be analogized to those with a hidden-strangeness pair~\cite{Lebed:2015dca, He:2017aps, Huang:2018ehi}, allowing us to propose a series of partners to the $P_c$ states in the strange sector—denoted here as $K\Sigma^\ast$, $K^\ast\Sigma$, and $K^\ast\Sigma^\ast$ bound states, namely $N(1875)$, $N(2080)$, and $N(2270)$ respectively—in the strange quark sector. The decay properties of these states have also been systematically investigated~\cite{Lin:2018kcc, Ben:2024qeg}. Furthermore, these states have also been employed with success in explaining the mechanisms of photoproduction reactions in $\gamma p \rightarrow \phi p$, $K^\ast\Sigma$, $K\Sigma$, and $K^\ast\Lambda$~\cite{Wu:2023ywu, Ben:2023uev, Suo:2025rty, Tian:2025bkx}.

Current understanding of the baryons remains limited. While the nucleon spectrum has been relatively well studied, gaps persist for both experimentally and theoretically in the spectra of $\Sigma$, $\Lambda$, and especially $\Xi$ baryons. To fill this blank, a naturally raised thought is that if the nucleon has its own octet partners, might these $N$ states contain a pair of $s\Bar{s}$ quarks that also possess corresponding multiplet partners in theory. These multiplet partners can provide crucial insights for advancing our understanding of baryonic spectroscopy, which can be further investigated in $\Bar{\nu}N$ and $\Bar{K}N$ scattering processes, such as $\Bar{K}N\rightarrow K\Xi$\cite{Burgun:1968ice,Carlson:1973td,Dauber:1969hg,Birmingham-Glasgow-LondonIC-Oxford-Rutherford:1966onr,London:1966zz,Trippe:1967wat,Trower:1968zz}.

To explore such possible multiplet partners, one of the most powerful and widely used tools in hadron physics-QCD sum rules~\cite{Shifman:1978bx,Shifman:1978by} is employed. Over the past four decades, this approach has played a crucial role in advancing our understanding of hadron spectroscopy~\cite{Agaev:2016mjb,Zhang:2025fuz,Albuquerque:2021tqd,Albuquerque:2020ugi,Wang:2013daa,Zhang:2022obn,Wan:2020oxt,Wan:2020fsk,Wang:2013vex,Tang:2019nwv,Wang:2015epa,Zhang:2024ulk,Wan:2022xkx,Qiao:2013dda,Qiao:2014vva,Zhang:2025vqg} and decays~\cite{Tan:2023opd,Aliev:2018vye,Zhang:2024asb,Azizi:2020tgh,Zhang:2023nxl,Dosch:1997zx,Huang:1998rq,Zhao:2020mod,Zhang:2024ick,Aliev:2010uy}. The basic principles and applications of QCD sum rules can be referred to in these reviews~\cite{Colangelo:2000dp,Nielsen:2009uh,Albuquerque:2018jkn,Wang:2025sic}.

In particular, for the QCD sum rules studies of hidden-charm pentaquark states, Chen et al. investigated the exotic hidden-charm pentaquarks $P_c(4380)$ and $P_c(4450)$ composed of an anticharmed meson and a charmed baryon, where the results indicate that $P_c(4380)$ and $P_c(4450)$ can be interpreted as $\bar{D}^*\Sigma_c$ with $J^P=3/2^-$ and mixed $\bar{D}\Sigma_c^*$-$\bar{D}^*\Lambda_c$ configurations with $J^P=5/2^+$~\cite{Chen:2015moa}, respectively. Wang also constructed the interpolating currents within the diquark-diquark-antiquark picture and the studies also supported to assign $P_c(4380)$ and $P_c(4450)$ as $J^P=3/2^-$ and $J^P=5/2^+$~\cite{Wang:2015epa}. In Ref.~\cite{Wang:2020eep}, the authors studied the scalar-diquark-scalar-diquark-antiquark type $u d s c \bar{c}$ for $P_{cs}(4459)$, where the analysis supported assigning this state with $J^P=1/2^-$. Other QCD sum rules studies for hidden-charm pentaquarks can be found in Refs.~\cite{Wang:2015wsa,Yang:2022uot,Chen:2019bip,Wang:2015ava,Chen:2016otp,Xiang:2017byz,Azizi:2016dhy}.

Thus, by following the ideas above, we systematically construct hidden- and open-strange hadronic molecules from the baryon octet combining with a pair of $s\Bar{s}$ quarks and discuss the spectrum obtained from QCD sum rules.

This paper is organized as follows. In Sec.~\ref{Sec:forma}, we briefly introduce the framework of our theoretical model. In Sec.~\ref{Sec:results}, the results of our theoretical calculations with some discussions are presented. Finally, in Sec.~\ref{sec:summary}, we give a brief summary and conclusions.

\section{Formalism}\label{Sec:forma}

In this work, the pentaquark currents constructed within the hadronic molecular picture can be interpreted at the quark level as being formed by coupling the constituent quarks of an octet baryon with an $s\Bar{s}$ pair. This pairing process involves two distinct configurations: (a) the strange and antistrange quarks form a mesonic component within the composite system; (b) the strange quark from the pair is incorporated into the baryonic component, while the antistrange quark enters the mesonic component. The corresponding quark compositions for these two configurations are summarized in Table~\ref{tab_comp}, where we use constituent hadrons to refer to the specific quark configurations and ignore the isospin differences.

Note that in case (b), where the strange quark is incorporated into the baryonic component, an $\Omega$ baryon will appear. However, the $\Omega$ belongs to the baryon decuplet and involves a spin parity $J^P = 3/2^+$ current. When the baryon decuplet is systematically taken into account, the situation becomes more interesting but also more complex, as there are more varieties of current combinations for each spin parity.
For simplicity, the baryonic currents are restricted with $J^P = 1/2^+$ in this work; therefore, the $\Omega$ baryon is not under consideration. Two types of mesonic component are examined, which are pseudoscalar octet ($J^P = 0^-$ current) and vector octet ($J^P = 1^-$ current). In Table~\ref{tab_comp}, these correspond to the particles shown in the right and left parentheses, respectively.

It should be mentioned that the quark-flavor scheme is one of the commonly used scheme of $\eta$ and $\eta'$ mixing~\cite{Feldmann:2002kz,Feldmann:1999uf}. In this scheme, the assumed orthogonal state $\eta_s$ are employed to incorporate a pseudoscalar current with a pure $s\Bar{s}$ in our construction. Using the mass $m_{ss}$ defined in the $\eta-\eta'$ quark-flavor mixing scheme~\cite{Feldmann:2002kz},
\begin{equation}
m_{ss}^2 = 2M_K^2-M_\pi^2,\label{eq_ss}
\end{equation}
we have obtained a mass value of $m_{\eta_s}=m_{ss}=683$ MeV. Thus, in the following discussion, we adopt $\eta_s$ to label this pseudoscalar current and to indicate its pure strange quark configuration.

\begin{table}[ht]
\renewcommand{\arraystretch}{1.5}
\centering
\caption{Calculated channels. Note that the constituent hadrons refer to the specific quark configurations}
\begin{tabular}{c|cccc}
Case (a)& $N(\phi/\eta_s)$ & $\Lambda(\phi/\eta_s)$ & $\Sigma(\phi/\eta_s)$ & $\Xi(\phi/\eta_s)$\\\hline
Case (b)& $-$ & $\Lambda(K^*/K)$ & $\Sigma (K^*/K)$ & $\Xi (K^*/K)$
\end{tabular}
\label{tab_comp}
\end{table}

In order to study the strange pentaquark states listed in Table~\ref{tab_comp} using QCD sum rules, the first step is to construct the corresponding currents. In our calculations, the interpolating currents for the constituent baryons and mesons are as follows:
\begin{align}
j_p &= \epsilon^{abc} (u_a^T C \gamma_5 d_b) u_c\,, \\[3pt]
j_{\Lambda} &= \epsilon^{abc} (u_a^T C \gamma_5 d_b) s_c\,, \\[3pt]
j_{\Sigma^+} &= \epsilon^{abc} (u_a^T C \gamma_5 s_b) u_c\,, \\[3pt]
j_{\Xi^0} &= \epsilon^{abc} (s_a^T C \gamma_5 u_b) s_c\,,\\[3pt]
j_\phi &=\Bar{s}_d\gamma^\mu s_d\,,\\[3pt]
j_{\eta_s} &=\Bar{s}_d\gamma_5 s_d\,,\\[3pt]
j_{K^{*+}} &=\Bar{s}_d\gamma^\mu u_d\,\\[3pt]
j_{K^+} &=\Bar{s}_d\gamma_5 u_d\,,
\end{align}
where $a$, $b$, $c$, and $d$ are color indices, and $C$ is the charge conjugation operator.

Within the hadronic molecular picture, we classify the spin parity of the pentaquark states based on the assumption that they are composed of constituent hadrons in an $s$-wave coupling (orbital angular momentum $L=0$). For a system consisting of a baryon from the octet and a pseudoscalar meson, the resulting spin parity is uniquely determined to be $J^P = 1/2^-$. For a system composed of a baryon octet and a vector meson, we consider both possible $s$-wave molecular configurations, which yield states with spin parity $J^P = 1/2^-$ and $J^P = 3/2^-$, respectively.

To construct the pentaquark currents described above, it is necessary to introduce appropriate Lorentz structures that endow the currents with the desired spin parity quantum numbers. For each case, the explicit form of the resulting pentaquark current for pseudoscalar meson case, after incorporating these Lorentz structures, is given by:
\begin{equation}
j_{1/2^-}=j_B j_P\,,
\end{equation}
where $j_B$ is the corresponding baryon current while the $j_P$ is the pseudoscalar meson current. The vector meson case is given by:
\begin{equation}
j_{1/2^-}=j_B\gamma_5\gamma_\mu j_V^\mu\,,
\end{equation}
\begin{equation}
j_{3/2^-}^\mu =j_B j_V^\mu\,,
\end{equation}
where $j_V^\mu$ is the vector meson current.

Thus, through the construction scheme outlined above and based on the constituent-hadron pairs listed in Table~\ref{tab_comp} along with the corresponding initial spin parity, a total of 21 distinct particle states have been systematically considered and calculated in our work.

After constructing the appropriate interpolating currents, the next step is to calculate the two-point correlation functions:
\begin{align}
\Pi(q^2) = i\int d^4x \,e^{iq\cdot x} \langle 0|T[j(x)\,,j^\dagger(0)]|0\rangle\,,\\
\Pi_{\mu\nu}(q^2) = i\int d^4x \,e^{iq\cdot x} \langle 0|T[j_\mu(x)\,,j_\nu^\dagger(0)]|0\rangle\,.
\end{align}
Here, $j(x)$ and $j_\mu(x)$ represent the interpolating currents for states with spin parity $J^P = 1/2^-$ and $J^P = 3/2^-$, respectively. The correlation functions can be decomposed into two independent Lorentz structures:
\begin{align}
\Pi(q^2) &= \slashed{q}\Pi_1(q^2) + \Pi_2(q^2)\,,\nonumber\\[5pt]
\Pi_{\mu\nu}(q^2) &= (-g_{\mu\nu} + \frac{q_\mu q_\nu}{q^2})(\slashed{q}\Pi_1(q^2)+\Pi_2(q^2))\cdots\,.   
\label{invariant-functions}                
\end{align}
For the spin-$3/2$ case, we retain only the Lorentz structure proportional to $(-g_{\mu\nu}+q_\mu q_\nu/q^2)$ in Eq.~\eqref{invariant-functions}. This structure projects out the pure spin-$3/2$ component, while the spin-$1/2$ contaminations are contained in the other tensor structures and are therefore not used in the present analysis. Two independent QCD sum rules can be derived for each invariant function $\Pi_1(q^2)$ and $\Pi_2(q^2)$. These two different Lorentz structure is decomposed to be a chiral-even structure $\Pi_1$ and a chiral-odd structure $\Pi_2$. These two pieces have different operator product expansion (OPE) content and probe different aspects of QCD dynamics. Practically, $\Pi_1$ often depends more on the perturbative and higher-dimension terms, while $\Pi_2$ is dominated by low-dimension condensates.

The correlation functions can be described at both the quark-gluon level and the hadron level, where the former is called QCD representation and the latter is called phenomenological representation. At the quark-gluon level, the correlation functions can be represented by the operator product expansion (OPE) and dispersion relations in the following form:

\begin{align}
\Pi^{\text{QCD}}(q^2) = \int_{s_{\min}}^\infty \frac{\rho^{\text{QCD}}(s)}{s-q^2}ds = \int_{s_{\min}}^\infty \frac{\slashed{q}\rho^{\text{QCD}}_1(s)+\rho^{\text{QCD}}_2(s)}{s-q^2}ds \,,
\end{align}
where $s_{\min}$ is the kinematic limit, and $i=1, 2$ denotes the two Lorentz structures presented in Eq.~\eqref{invariant-functions}. $\rho^{\text{QCD}}_i(s)$ is the spectral density, which is given by the imaginary part of the correlation function:
\begin{align}
\rho^{\text{QCD}}_i(s) = \frac{1}{\pi} \,\text{Im}\,\Pi^{\text{QCD}}_i(s)\,.
\end{align}
In this work, contributions up to dimension 11 are taken into account in the OPE series, which can be expressed as:
\begin{align}
\rho_i^{\text{QCD}}(s)=&\rho_i^{\text{pert}}(s)+\rho_i^{\langle\Bar{q}q\rangle}(s)+\rho_i^{\langle g_s^2 G^2\rangle}(s)+\rho_i^{\langle g_s \Bar{q}\sigma\cdot G q\rangle}(s)+\rho_i^{\langle\Bar{q}q\rangle^2}(s)+\rho_i^{\langle g_s^3 G^3\rangle}(s)+\rho_i^{\langle\Bar{q}q\rangle\langle g_s^2 G^2\rangle}(s)\nonumber\\[3pt]
+&\rho_i^{\langle\Bar{q}q\rangle\langle g_s \Bar{q}\sigma\cdot G q\rangle}(s)+\rho_i^{\langle g_s^2 G^2\rangle^2}(s)+\rho_i^{\langle g_s^2 G^2\rangle\langle g_s \Bar{q}\sigma\cdot G q\rangle}(s)+\rho_i^{\langle\Bar{q}q\rangle^3}(s)+\rho_i^{\langle g_s \Bar{q}\sigma\cdot G q\rangle^2}(s)\nonumber\\
+&\rho_i^{\langle\Bar{q}q\rangle^2\langle g_s^2 G^2\rangle}(s)+\rho_i^{\langle\Bar{q}q\rangle^2\langle g_s \Bar{q}\sigma\cdot G q\rangle}(s)\,.
\end{align}

For the higher-dimensional condensates in the OPE, a nontrivial choice of factorization scheme is required~\cite{Launer:1983ib,Reinders:1984sr,Narison:1989aq,Narison:2009vy}. The possible deviations from the factorization hypothesis are parametrized by a coefficient $\kappa$, for which $\kappa=1$ corresponds to exact vacuum saturation. A unified parametrization of $\kappa$ is introduced for all factorized condensates according to Ref.~\cite{Gubler:2009iq}:
\begin{align}
\langle\bar{q}q\,\bar{q}q\rangle &= \kappa\,\langle\bar{q}q\rangle^2, &
\langle\bar{q}q\,\bar{q}q\,\bar{q}q\rangle &= \kappa^2\,\langle\bar{q}q\rangle^3, \nonumber\\
\langle\bar{q}q\,\bar{q}\,g\sigma\!\cdot\!G\,q\rangle &= \kappa\,\langle\bar{q}q\rangle\langle\bar{q}\,g\sigma\!\cdot\!G\,q\rangle, &
\langle\bar{q}\,g\sigma\!\cdot\!G\,q\,\bar{q}\,g\sigma\!\cdot\!G\,q\rangle &= \kappa\,\langle\bar{q}\,g\sigma\!\cdot\!G\,q\rangle^2,\;\ldots
\end{align}
The vacuum saturation approximation has been shown to be valid in the leading order of the large-$N_c$ expansion, even though the $1/N_c$ corrections may be quite large~\cite{Gubler:2009iq}. In this work, $\kappa$ is varied over the range $1$--$2.1$~\cite{Albuquerque:2018jkn,Nielsen:2009uh}, and the resulting uncertainty is propagated into the final results.

To evaluate the spectral density of the QCD representation, the full propagators for light quarks ($u$, $d$, and $s$) are given by~\cite{Reinders:1984sr}:
\begin{align}
S_{ab}^q(x) &= \langle 0|T[q_a(x)\Bar{q}_b(0)]|0\rangle = \frac{i\delta_{ab}\slashed{x}}{2\pi^2 x^4 } - \frac{m_q\delta_{ab}}{4\pi^2 x^2}- \frac{\delta_{ab}}{12}\langle\Bar{q}q\rangle + \frac{i\delta_{ab}\slashed{x}}{48}m_q\langle\Bar{q}q\rangle- \frac{\delta_{ab}x^2}{192}\langle g_s\Bar{q}\sigma\cdot G q\rangle  \nonumber\\[5pt]
&- \frac{ig_s t^A_{ab} G^A_{\mu\nu}}{32\pi^2 x^2}(\slashed{x}\sigma^{\mu\nu}+ \sigma^{\mu\nu}\slashed{x}) + \frac{i\delta_{ab}x^2 \slashed{x}}{1152}m_q\langle g_s\Bar{q}\sigma\cdot G q\rangle- \frac{t^A_{ab}\sigma^{\mu\nu}}{192}\langle g_s\Bar{q}\sigma\cdot G q\rangle +\cdots\,,
\end{align}
where $a$ and $b$ are color indices, $m_q$ is the quark mass, $t^A = \lambda^A/2$ with $\lambda^A$ being the Gell-Mann matrices, and $G^A_{\mu\nu}$ is the gluon field strength tensor. The first two terms of the full propagator represent the perturbative contributions, while the remaining terms account for nonperturbative effects, which are parametrized by various condensates.

At the hadron level, the correlation functions can be expressed in terms of hadronic observables using dispersion relations:

\begin{align}
\Pi^{\text{phe}}(s) &= \lambda_-^2(\slashed{q}+M_-)\delta(s-M_-^2)+\lambda_+^2(\slashed{q}-M_+)\delta(s-M_+^2)+\cdot\cdot\cdot\nonumber\\
&=\slashed{q}\big(\lambda_-^2\delta(s-M_-^2)+\lambda_+^2\delta(s-M_+^2)\big)+\big(\lambda_-^2 M_-\delta(s-M_-^2)-\lambda_+^2 M_+\delta(s-M_+^2)\big)+\cdot\cdot\cdot\nonumber\\
&=\slashed{q}\rho_1^{\text{phe}}(s)+\rho_2^{\text{phe}}(s)\,,
\end{align}
where $M_\pm$ is the mass of the hadron and $\lambda_\pm$ is the coupling constant between the current and the hadron. After constructing parity-projected spectral densities $\rho_\pm^{\text{phe}}(s)=\sqrt{s}\,\rho_1^{\text{phe}}(s)\mp\rho_2^{\text{phe}}(s)$, one can obtain:
\begin{align}
\rho_-^{\text{phe}}(s)
&=\sqrt{s}\big(\lambda_-^2\delta(s-M_-^2)+\lambda_+^2\delta(s-M_+^2)\big)
-\big(\lambda_-^2 M_-\delta(s-M_-^2)-\lambda_+^2 M_+\delta(s-M_+^2)\big)\nonumber\\
&=2M_+\lambda_+^2\delta(s-M_+^2)\,,
\end{align}
and
\begin{align}
\rho_+^{\text{phe}}(s)
&=\sqrt{s}\big(\lambda_-^2\delta(s-M_-^2)+\lambda_+^2\delta(s-M_+^2)\big)
+\big(\lambda_-^2 M_-\delta(s-M_-^2)-\lambda_+^2 M_+\delta(s-M_+^2)\big)\nonumber\\
&=2M_-\lambda_-^2\delta(s-M_-^2)\,.
\end{align}

In the present analysis, we employ the zero-width approximation, i.e., the resonance contribution to the spectral density is represented by a $\delta$ function. Finite-width effects will shift the pole position; however, the mass extraction is rather insensitive to the width variation, and the mass in the zero-width approximation agrees with that including finite-width effects within their uncertainties, as demonstrated for $\Delta(1232)$~\cite{Erkol:2008gp}. Furthermore, the simultaneous fit of mass and width has been shown to be not feasible in parity-projected sum rules~\cite{Erkol:2008gp,Ohtani:2012ps}, the same framework employed here. For the pentaquark configurations constructed in this work, reliably determining the decay widths is itself a challenge, as a reasonable estimate of the total width would require more involved calculations such as triangle-loop diagrams\cite{Lin:2018kcc, Ben:2024qeg}. As this work focuses on the mass spectrum, the total width is left for future studies.

After performing the Borel transformation and invoking quark-hadron duality, the sum rules can be written as
\begin{align}
2M_\pm \lambda_\pm^2 e^{-M_\pm^2/M_B^2}
=\int_{s_{\min}}^{s_0} ds\,\Big[\sqrt{s}\,\rho_1^{\text{QCD}}(s)\mp\rho_2^{\text{QCD}}(s)\Big]e^{-s/M_B^2}\,,
\end{align}
from which the hadron masses of definite parity can be extracted as
\begin{align}
M_{H,\pm}^2=
\frac{\int_{s_{\min}}^{s_0} ds\, s\Big[\sqrt{s}\,\rho_1^{\text{QCD}}(s)\mp\rho_2^{\text{QCD}}(s)\Big]e^{-s/M_B^2}}
{\int_{s_{\min}}^{s_0} ds\, \Big[\sqrt{s}\,\rho_1^{\text{QCD}}(s)\mp\rho_2^{\text{QCD}}(s)\Big]e^{-s/M_B^2}}\,.
\end{align}
Here, $s_0$ is the threshold parameter that separates the ground state from higher excited states and continuum contributions. {For comparison, in numerical analysis we present both the parity-projected results and the results obtained separately from the chiral-even and chiral-odd structures from Eq.~\eqref{invariant-functions}.}

The analytical results for the spectral densities are lengthy, and, thus, we showcase only the expression of $p\eta^\prime (1/2^-)$:
\begin{align}
\rho_1^{\text{QCD}}(s) &= -\frac{s^5}{137625600\pi^8}-\frac{m_s^2\,s^4}{3932160\pi^8}-\frac{\langle g_s^2 G^2\rangle s^3}{5242880\pi^8}+\frac{m_s \langle \bar{s}s\rangle s^3}{122880\pi^6}+\frac{m_s^2 \langle g_s^2 G^2\rangle s^2}{1048576\pi^8}\nonumber\\[5pt]
&-\frac{m_s \langle g_s\Bar{s}\sigma\cdot G s\rangle s^2}{24576\pi^6}
-\frac{\langle \bar{q}q\rangle^2 s^2}{6144\pi^4}-\frac{\langle \bar{s}s\rangle^2 s^2}{6144\pi^4}+\frac{\langle \bar{s}s\rangle \langle g_s\Bar{s}\sigma\cdot G s\rangle s}{1536\pi^4}+\frac{\langle \bar{q}q\rangle \langle g_s\Bar{q}\sigma\cdot G q\rangle s}{1536\pi^4}\nonumber\\[5pt]
&+\frac{\langle g_s^2 G^2\rangle^2 s}{3145728\pi^8}-\frac{m_s^2 \langle \bar{q} q \rangle^2 s}{768\pi^4}-\frac{m_s \langle \bar{s}s\rangle \langle g_s^2 G^2\rangle s}{98304\pi^6}-\frac{m_s^2 \langle \bar{s}s\rangle^2 s}{3072\pi^4}+\frac{m_s \langle \bar{q}q\rangle^2 \langle \bar{s}s\rangle}{192\pi^2}\nonumber\\[5pt]
&+\frac{m_s^2\langle \bar{q}q\rangle \langle g_s\Bar{q}\sigma\cdot G q\rangle }{512\pi^4}+\frac{m_s^2 \langle \bar{s}s\rangle \langle g_s\Bar{s}\sigma\cdot G s\rangle}{3072\pi^4}+\frac{m_s\langle g_s^2 G^2\rangle  \langle g_s\Bar{s}\sigma\cdot G s\rangle}{98304\pi^6}-\frac{\langle \bar{q}q\rangle^2 \langle g_s^2 G^2\rangle}{6144\pi^4}\nonumber\\[5pt]
&+\frac{\langle g_s^2 G^2\rangle \langle \bar{s}s\rangle^2}{24576\pi^4}-\frac{\langle g_s\Bar{q}\sigma\cdot G q\rangle^2}{4096\pi^4}-\frac{\langle g_s\Bar{s}\sigma\cdot G s\rangle^2}{4096\pi^4}\,,
\end{align}
\begin{align}
\rho_2^{\text{QCD}}(s) &=\frac{\langle \bar{q}q\rangle s^4}{491520\pi^6}+\frac{m_s^2\langle \bar{q}q\rangle s^3}{24576\pi^6}-\frac{\langle g_s\Bar{s}\sigma\cdot G s\rangle s^3}{98304\pi^6}+\frac{\langle \bar{q}q\rangle \langle g_s^2 G^2\rangle s^2}{65536\pi^6}-\frac{m_s^2\langle g_s\Bar{s}\sigma\cdot G s\rangle s^2}{8192\pi^6}\nonumber\\[5pt]
&-\frac{m_s\langle \bar{q}q\rangle \langle \bar{s}s\rangle s^2}{1536\pi^4}-\frac{m_s^2\langle \bar{q}q \rangle \langle g_s G^2\rangle s}{32768\pi^6}+\frac{m_s \langle \bar{q}q\rangle \langle g_s\Bar{s}\sigma\cdot G s\rangle s}{768\pi^4}+\frac{m_s \langle g_s\Bar{q}\sigma\cdot G q\rangle \langle \bar{s}s\rangle s}{1024\pi^4}\nonumber\\[5pt]
&+\frac{\langle \bar{q}q\rangle^3 s}{192\pi^2}+\frac{\langle \bar{q}q\rangle \langle \bar{s}s\rangle^2 s}{192\pi^2}-\frac{3 \langle g_s G^2\rangle \langle g_s\Bar{q}\sigma\cdot G q\rangle s}{131072\pi^6}\,,
\end{align}
\begin{align}
\Pi^{sum}(s)&=\frac{\slashed{q}}{s} \big(\frac{m_s\langle \bar{q}q \rangle^2  \langle g_s\Bar{s}\sigma\cdot G s \rangle}{288\pi^2}-\frac{m_s^2\langle g_s^2 G^2 \rangle  \langle \bar{s}s \rangle^2}{49152\pi^4}+\frac{m_s\langle \bar{q}q \rangle \langle g_s\Bar{q}\sigma\cdot G q \rangle  \langle \bar{s}s \rangle}{192\pi^2}+\frac{m_s^2\langle g_s\Bar{q}\sigma\cdot G q \rangle^2 }{2048\pi^4}\nonumber\\[5pt]
&+\frac{m_s^2 \langle g_s\Bar{s}\sigma\cdot G s \rangle^2}{18432\pi^4}\big)\nonumber\\[5pt]
&+\frac{1}{s}\big(\frac{m_s^2\langle \bar{q}q \rangle^2 \langle g_s\Bar{q}\sigma\cdot G q \rangle }{128\pi^2}+\frac{m_s^2\langle g_s\Bar{q}\sigma\cdot G q \rangle  \langle \bar{s}s \rangle^2}{1536\pi^2}+\frac{m_s^2\langle \bar{q}q \rangle  \langle \bar{s}s \rangle \langle g_s\Bar{s}\sigma\cdot G s \rangle}{1152\pi^2}  \big)\,,
\end{align}
where $\Pi^{sum}(s)$ collects terms of the correlation function with vanishing imaginary part, which nevertheless produce nontrivial effects upon Borel transformation. We observe that, for certain states like $p\eta^\prime (1/2^-)$, the spectral density $\rho_2^{\text{QCD}}(s)$ contains no perturbative term, which is consistent with Ref.~\cite{Yang:2022uot}.

\section{Results and Discussion}\label{Sec:results}

In the numerical analysis, the following inputs are adopted~\cite{ParticleDataGroup:2024cfk,Colangelo:2000dp,Wan:2021vny,Reinders:1984sr,Ioffe:1981kw,Reinders:1984sr}:
\begin{align}
&\langle \bar{q}q \rangle = -(0.24 \pm 0.01)^3 \ \text{GeV}^3\,, 
\langle \bar{s}s \rangle = (1.15 \pm 0.12)\langle \bar{q}q \rangle\,, \nonumber\\[3pt]
&\langle \bar{q}g_s \sigma \cdot G q \rangle = m_0^2 \langle \bar{q}q \rangle\,, 
\langle \bar{s}g_s \sigma \cdot G s \rangle = m_0^2 \langle \bar{s}s \rangle\,, \nonumber\\[3pt]
&\langle g_s^2 G^2 \rangle = (0.88 \pm 0.25) \ \text{GeV}^4\,,\langle g_s^3 G^3 \rangle = (0.045 \pm 0.013) \ \text{GeV}^6\,,\nonumber\\[3pt]
&m_0^2 = (0.8 \pm 0.1) \ \text{GeV}^2\,, m_s = (95 \pm 5) \ \text{MeV}\,.
\end{align}
 The quark masses are given in the $\overline{\mathrm{MS}}$ scheme, and the condensates correspond to a common renormalization scale $\mu\sim 1$\,GeV~\cite{Zhang:2025qmg, Colangelo:2000dp,Reinders:1984sr}. In addition to the above QCD parameters, the variation of the factorization-breaking parameters in the range $\kappa=1$--$2.1$ around the vacuum-saturation limit is also considered. Moreover, the masses of the $u$ and $d$ quarks are neglected.

QCD sum rules involve two crucial parameters, which are the Borel parameter $M_B^2$ and the threshold parameter $s_0$. The selection of appropriate ranges for these two parameters is essential to ensure the reliability of the QCD sum rule analysis. The determination of these two parameters is typically guided by two main criteria. First, the convergence of the OPE series must be ensured, which requires that the contributions from the highest-dimensional condensates are sufficiently suppressed. In this work, to guarantee convergence of the OPE we demand that the contribution from the highest-dimension term remain below 5\% of the total OPE contribution, i.e.
\begin{equation}
R^{\text{OPE}} = \frac{\int_{s_{\min}}^{s_0} ds \,\rho_{1,D_{max}}^{\text{QCD}}}{\int_{s_{\min}}^{s_0} ds \,\rho_1^{\text{QCD}}} < 5\%\,,
\end{equation}
where $\rho_{1,D_{max}}^{\text{QCD}}$ represents the contribution from the highest-dimensional condensate considered in the OPE series.

Second, the pole contribution (PC) from the ground state should be substantial compared to the total contribution. As noted in Refs.~\cite{Chen:2014vha,Azizi:2019xla,Wang:2017sto}, the high power of $s$ in the spectral density will suppress the PC value; accordingly, the pole contribution will be required larger than 20\%, i.e.
\begin{align}
R^{\text{PC}} = \frac{\int_{s_{\min}}^{s_0} ds \,\rho_1^{\text{QCD}}(s) e^{-s/M_B^2}}{\int_{s_{\min}}^{\infty} ds \,\rho_1^{\text{QCD}}(s) e^{-s/M_B^2}}>20\%\,.
\end{align}
Furthermore, since the hadron mass is a physical observable, it should be independent of auxiliary parameters. Consequently, a trustworthy prediction is obtained in an optimal window where the extracted mass is minimally sensitive to the Borel parameter $M_B^2$.

In practice, we gradually adjust $\sqrt{s_0}$ by 0.1 GeV within the range $\sqrt{s_0} \sim (E_{th}+\delta)$ to find the acceptable range of the Borel
parameters, where $E_{th}$ is the two-particle threshold of the constituent states. Since $s_0$ is the parameter that separates the ground state from the excited states and continuum, the choice of $\delta$ should, in principle, ensure that the extracted hadronic state corresponds to the ground state. In the heavy-quark sector, $\delta$ is generally taken to be less than or equal to 1 GeV~\cite{Wu:2021tzo,Wu:2022qwd,Wu:2023ntn}. However, in the strange regime relevant to our study, RPP(PDG)~\cite{ParticleDataGroup:2024cfk} provides substantially more spectroscopic information. The mass gap between the ground states and excited states for several baryons are compiled in Table~\ref{tab_mass}, where all observed mass splittings cluster around $0.5$ GeV, except for the $\Xi$ baryon whose excitation spectrum remains incomplete.

\begin{table}[ht]
\renewcommand{\arraystretch}{1.5}
\centering
\caption{Mass statistics listed on RPP(PDG)~\cite{ParticleDataGroup:2024cfk} of ground states and their first excitation within the same quantum number, where $m_G$ is the mass of the ground state, $m_E$ is the mass of the lowest excited state, and $\delta_m$ is the mass difference. We truncated the numerical values to the nearest integer.}
\begin{tabular}{c|c|c|c}
Particle & $m_G$ (MeV) & $m_E$ (MeV) & $\delta_m$ (MeV)\\\hline
$N$ & $939$ & $1440$ & $501$\\
$\Lambda$ & $1116$ & $1600$ & $484$\\
$\Sigma$ & $1193$ & $1660$ & $467$\\
$\Xi$ & $1318$ & $\cdots$ & $\cdots$\\
\end{tabular}
\label{tab_mass}
\end{table}

Based on this observation, the following criterion is adopted in our analysis: If a stable Borel window and a corresponding hadron state can be identified with $ \delta \leq 0.5\, \text{GeV} $, we interpret the state as the ground state associated with the constructed current. Otherwise, the state is regarded as an excited state contribution. The latter case also implies that the chosen current may not be adequate for describing the true ground state in that channel.

Through the above procedure, the variations of the pole contribution, OPE convergence, and the mass with respect to the Borel parameter $M_B^2$ are presented in the Appendix. The numerical results are listed in Tables~\ref{tab_result_Pminus}, \ref{tab_result_Pplus}, and \ref{tab_result_V3}, with all uncertainties from the QCD parameters, Borel parameters, threshold parameters, and the variation of $\kappa$ taken into account. We find that the contribution from the variation of $\kappa$, i.e., the uncertainty induced by the factorization approximation for higher-dimensional condensates, amounts to approximately $1\%$.

\begin{figure*}[ht]
\centering
\includegraphics[width=0.8\textwidth]{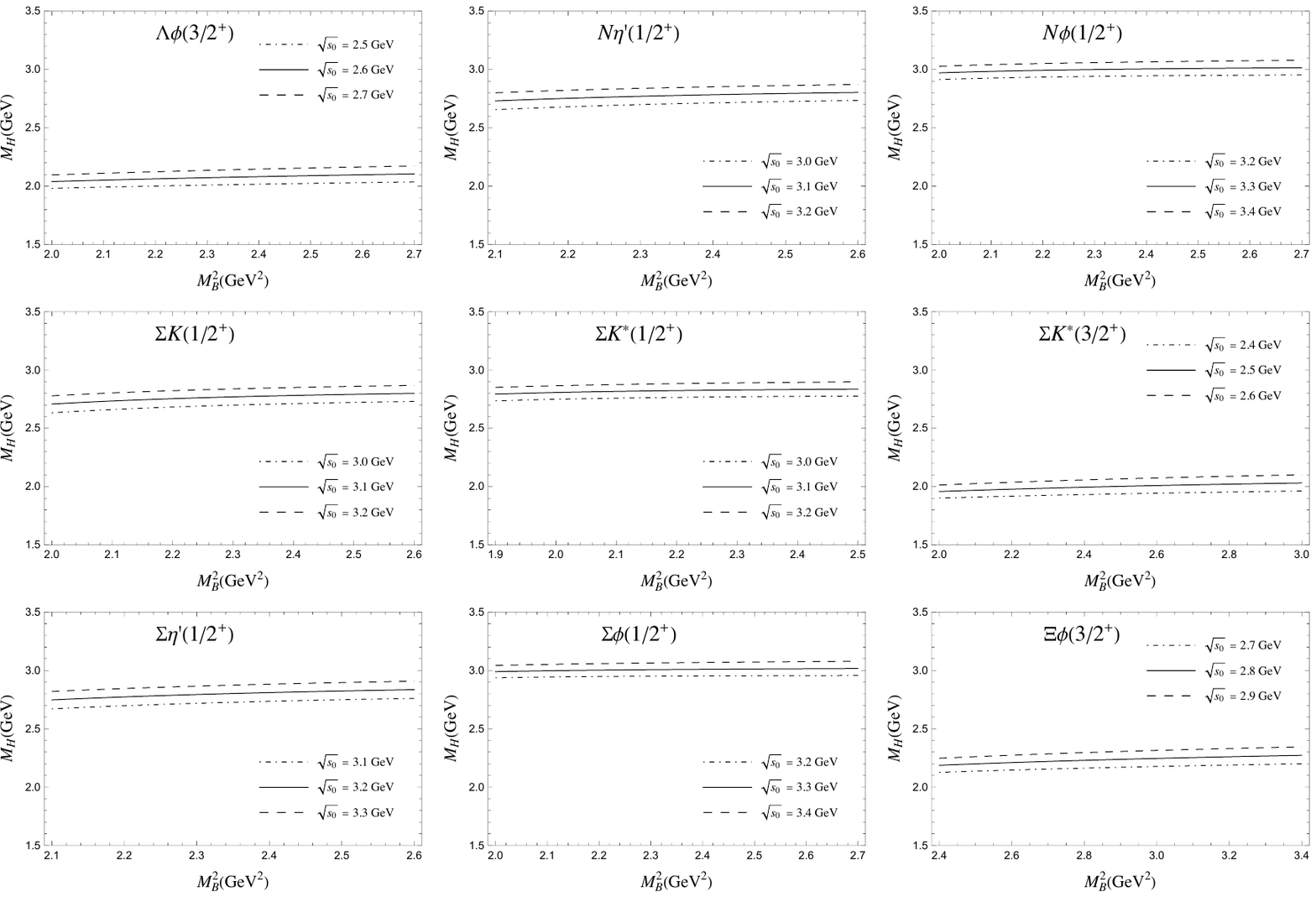}
\caption{The mass with respect to the Borel parameter $M_B^2$ with different threshold parameter $\sqrt{s_0}$ for $\Pi_{+}$.}
\label{fig:M-Piplus}
\end{figure*}

It should be noted that the phenomenological spectral density may also receive contributions from meson-baryon scattering states carrying the same quantum numbers as the interpolating current~\cite{Gubler:2009iq,Sugiyama:2003zk,Gubler:2009ay}. Such scattering states produce a spectral function that grows rapidly with energy, leading to a strong dependence of the extracted mass on the threshold parameter $s_0$, whereas a narrow resonance pole gives a mass weakly dependent on $s_0$~\cite{Gubler:2009iq}. As shown in Figs.~\ref{fig:M-Pi1}, \ref{fig:M-Pi2}, \ref{fig:M-Piplus}, and \ref{fig:M-Piminus}, the extracted masses in the Borel windows vary only weakly with $s_0$, indicating that a narrow pole rather than scattering states dominates the spectral function. A more rigorous treatment to further suppress scattering-state contamination, such as the projected correlation function approach~\cite{Gubler:2009iq, Kojo:2006bh}, would be valuable for confirmation.

In following discussion, we will focusing on the result of negative parity since our purpose is to investigate the possible molecules. Meanwhile, we calculated the positive-parity states to fill the full spectrum. In addition, to make strong interpretation, the mass spectrum generated from chiral-even and chiral-odd structure $\Pi_1$ and $\Pi_2$ are also provided. The discussion will follow the order of $J = 1/2$ baryon pseudoscalar meson system, $J = 1/2$ baryon vector meson system, and $J = 3/2$ baryon vector meson system.

\subsection{$1/2^+$ and $0^-$ to $J = 1/2$ system}

Since the currents we constructed provides only the assumption of quark component and spin parity, that means, for instance, $K$ and $K(1460)$ share same current due to the same quark configuration and spin parity $J^P = 0^-$. Thus, following the selection rules above, when the threshold parameter $\delta > 0.5$ GeV, we consider the scenario in which one or both of the two constituent hadrons may exist in their excited states within the composite system.

It is important to emphasize that we aim to search for molecular states containing a pair of strange and antistrange quarks ($s\bar s$). Based on the $S$-wave spin-orbital ($L-S$) coupling, where $L=0$, the overall parity of the required pentaquark state is determined to be $-1$. We first examine, within the framework of QCD sum rules, whether a bound state can be obtained for a system with spin parity $J^P = 1/2^-$.

As shown in Table~\ref{tab_result_Pminus}, we are unable to find any bound states in the $1/2^-$ sector. It is crucial to emphasize that the molecules and candidates listed in our table do not imply that the specific current we calculated corresponds directly to such molecular components or configurations. We simply select nearby states to serve as possible references. To provide a concrete example, we find that the resonance energy obtained from the $N\eta_s$ current calculation is near 2.4 GeV. The combination of $N$ or $N^*$ with $J^P=1/2^+$ and $\eta$ or $\eta^*$ with $J^P=0^-$ that is closest to this energy corresponds to the $N(1440)1/2^+$ and $\eta'$ combination. However, based on current understanding, $\eta$ and $\eta'$ mixing exists, which is also discussed in Eq.~\ref{eq_ss}, and neither can be fully explained as a pure $s\bar{s}$ pseudoscalar meson state in terms of quark configuration. Therefore, using a current with a pure $ss$ configuration to represent $\eta'$ would be biased. Consequently, the possible molecule combinations listed here serve merely as a potential reference. Furthermore, there is no direct connection between the molecules and candidates in tables, and the only link between them is that both the molecules and candidates happen to fall within the calculated resonance mass and its error margin. 

A more intriguing observation is that we find that for both the $\Lambda\eta_s$ and $\Sigma\eta_s$ currents, the obtained resonance masses fall precisely near the $\Lambda \eta(1295)$ and $\Sigma \eta(1295)$ thresholds, respectively. In terms of current construction, this conversely implies that the $\eta(1295)$ is likely dominated by an $s \bar{s}$ structure.

Furthermore, the result for the $\Xi \eta_s$ current falls precisely in the vicinity of $\Xi(2370)$, while the result for the $\Lambda K$ current is located near the $\Lambda K(1460)$ two-body threshold. Additionally, $\Sigma(2455)$ falls within the resonance mass range corresponding to both the $\Sigma \eta_s$ and $\Xi K$ currents. Although the specific numerical results of these two currents differ, their quark configurations are both $qqss\bar{s}$ from the perspective of quark content. We currently know little about $\Sigma(2455)$, and it remains a one-star resonance in the RPP(PDG). However, our calculation results suggest that $\Sigma(2455)$ may contain a pentaquark component with the $qqss\bar{s}$ configuration, and mixing may exist between the color-singlet cluster combinations of the aforementioned two currents. Meanwhile, the possible spin parity assignment of $\Sigma(2455)$ is $1/2^-$.

\begin{table*}[ht]
\renewcommand{\arraystretch}{1.5}
\caption{$J^P=1/2^-$ states generated from a baryon octet and a pseudoscalar meson. Where constituent hadrons refer to the specific quark configurations, $E_{th}$ is the two-particle threshold of constituent hadrons, $M^2_B$ is the energy range of the Borel window, $\sqrt{s_0}$ is the threshold parameter, $M_H$ is the calculated mass, molecule implies that the nearest possible two-particle threshold, and candidate is the possible state which listed on RPP(PDG)~\cite{ParticleDataGroup:2024cfk}.}
\centering
\resizebox{\textwidth}{!}{
\begin{tabular}{c|c|c|c|c|c|c}
Constituent Hadrons  & $E_{th}$ (GeV) & $M_B^2$ (GeV$^2$) & $\sqrt{s_0}$ (GeV) & $M_H$ (GeV)& Molecule & Candidate\\\hline
$N\eta_s$ & $1.621$ & $1.6$--$2.5$ & $2.7\pm0.1$ & $2.42\pm0.08$ & $N(1440)\eta'$ & $-$ \\
$\Lambda\eta_s$ & $1.798$ & $1.8$--$2.6$ & $2.8\pm0.1$ & $2.48\pm0.08$ & $\Lambda\eta(1295)$ & $-$\\
$\Sigma\eta_s$ & $1.899$ & $1.6$--$2.4$ & $2.7\pm0.1$ & $2.38\pm0.08$ & $\Sigma\eta(1295)$ &  $\Sigma(2455)$\\
$\Xi\eta_s$ & $1.998$ & $1.7$--$2.5$ & $2.7\pm0.1$ & $2.41\pm0.08$ & $-$ & $\Xi(2370)$\\
$\Lambda K$ & $1.609$ & $1.8$--$2.4$ & $2.8\pm0.1$ & $2.49\pm0.08$ & $\Lambda K(1460)$ & $-$ \\
$\Sigma K$ & $1.683$ & $1.8$--$2.4$ & $2.7\pm0.1$ & $2.42\pm0.08$ & $-$ & $-$ \\
$\Xi K$ & $1.808$ & $1.7$--$2.5$ & $2.8\pm0.1$ & $2.48\pm0.08$ &  $-$ & $\Sigma(2455)$
\end{tabular}
}
\label{tab_result_Pminus}
\end{table*}

It is well known that baryon currents can simultaneously project onto both positive- and negative-parity components. Therefore, following the calculation of the negative-parity sector $J^P = 1/2^-$, we can likewise obtain the results for the positive-parity part, which are listed in Table~\ref{tab_result_Pplus}.

It is worth mentioning that we did not introduce additional orbital angular momentum when constructing the currents. In other words, the coupling between the two color-singlet structures within the molecular currents we constructed remains in the form of $s$-wave coupling. However, when we project onto the $P=+1$ sector, we recognize that, while the parity of the meson current component is fixed, the baryon current simultaneously contains both positive- and negative-parity components. Within such a framework, a possible interpretation in the molecular picture is that the positive-parity part calculated here corresponds to a $J^P=1/2^-$ baryon $s$-wave coupled to a $J^P=0^-$ pseudoscalar meson, thereby yielding a $1/2^+$ pentaquark current structure.

\begin{table*}[ht]
\renewcommand{\arraystretch}{1.5}
\caption{Ground states of $J^P=1/2^-$ within QCD sum rule\cite{Oka:1996zz}.}
\centering{
\begin{tabular}{c|c|c|c|c}
Baryon  & $N_-$ & $\Lambda_-$ & $\Sigma_-$ & $\Xi_-$\\\hline
Mass (GeV) & $1.54$ & $1.55$ & $1.63$ & $1.63$
\end{tabular}
}
\label{tab_minus_mass}
\end{table*}

Thus, to use actual physical designation refers to a three-quark color-singlet structure with spin parity $1/2^-$, we select $1/2^-$ states based on the calculation within QCD sum rule framework\cite{Oka:1996zz}, listed in Table~\ref{tab_minus_mass} and named as $N_-$, $\Lambda_-$, $\Sigma_-$, and $\Xi_-$, as the ground states to determine our corresponding threshold values and constrain the threshold parameter $\sqrt{s_0}$. 

It is interesting that, when we project onto the positive-parity sector, we find that it is extremely difficult to identify a corresponding state for the currents we constructed. Only $N_-\eta_s$, $\Sigma_-\eta_s$, and $\Sigma_-K$ find resonances at very high energy which are above $2.7$ GeV.

\begin{table*}[ht]
\renewcommand{\arraystretch}{1.5}
\caption{$J^P=1/2^+$ states generated from a baryon octet and a pseudoscalar meson. The notation is the same as its in Table~\ref{tab_result_Pminus}.}
\centering
\resizebox{\textwidth}{!}{
\begin{tabular}{c|c|c|c|c|c|c}
Constituent hadrons  & $E_{th}$ (GeV) & $M_B^2$ (GeV$^2$) & $\sqrt{s_0}$ (GeV) & $M_H$ (GeV)& Molecule & Candidate\\\hline
$N_-\eta_s$ & $2.223$ & $2.1$--$2.6$ & $3.1\pm 0.1$ & $2.78\pm 0.11$ & $N(1535)\eta(1295)$ & $-$ \\
$\Lambda_-\eta_s$ & $2.233$ & $\cdots$ & $\cdots$ & No state & $-$ & $-$\\
$\Sigma_-\eta_s$ & $2.313$ & $2.1$--$2.6$ & $3.2\pm0.1$ & $2.79\pm0.11$ & $\Sigma(1900)\eta'$ &  $-$\\
$\Xi_-\eta_s$ & $2.313$ & $\cdots$ & $\cdots$ & No state & $-$ & $-$\\
$\Lambda_- K$ & $2.045$ & $\cdots$ & $\cdots$ & No state & $-$ & $-$\\
$\Sigma_- K$ & $2.125$ & $2.0$--$2.6$ & $3.1\pm0.1$ & $2.76\pm0.10$ & $-$ & $-$\\
$\Xi_- K$ & $2.125$ & $\cdots$ & $\cdots$ & No state & $-$ & $-$
\end{tabular}
}
\label{tab_result_Pplus}
\end{table*}

\subsection{$1/2^+$ and $1^-$ to $J = 1/2$ system}

\begin{table*}[ht]
\renewcommand{\arraystretch}{1.5}
\caption{$J^P=1/2^-$ states generated from a baryon octet and a vector meson. The notation is the same as its in Table~\ref{tab_result_Pminus}.}
\centering
\resizebox{\textwidth}{!}{
\begin{tabular}{c|c|c|c|c|c|c}
Constituent hadrons  & $E_{th}$ (GeV) & $M_B^2$ (GeV$^2$) & $\sqrt{s_0}$ (GeV) & $M_H$ (GeV)& Molecule & Candidate\\\hline
$N\phi$ & $1.958$ & $2.0$--$2.8$ & $3.0\pm0.1$ & $2.67\pm0.08$ &  $N\phi(1680)$ & $-$\\
$\Lambda\phi$ & $2.135$ & $2.0$--$2.6$ & $3.0\pm0.1$ & $2.69\pm0.09$ & $\Lambda(1600)\phi$ & $-$\\
$\Sigma\phi$ & $2.209$ & $1.8$--$2.3$ & $2.8\pm0.1$ & $2.56\pm0.07$ & $-$ & $\Sigma(2620)$\\
$\Xi\phi$ & $2.334$ & $1.9$--$2.5$ & $2.9\pm0.1$ & $2.62\pm0.08$ & $-$ & $-$\\
$\Lambda K^\ast$ & $2.007$ & $2.0$--$2.6$ & $3.0\pm0.1$ & $2.70\pm0.08$ & $\Lambda(1810)K^\ast$ & $-$\\
$\Sigma K^\ast$ & $2.081$ & $1.9$--$2.5$ & $2.9\pm0.1$ & $2.64\pm0.08$ & $-$ & $-$\\
$\Xi K^\ast$ & $2.206$ & $2.0$--$2.5$ & $2.8\pm0.1$ & $2.59\pm0.07$ & $-$ & $\Sigma(2620)$
\end{tabular}
}
\label{tab_result_Vminus}
\end{table*}

When it comes to the $J^P=1/2$ sector generated from a baryon octet and a vector meson, the calculated mass of $J^P=1/2^-$ states are listed in Table~\ref{tab_result_Vminus}. Still, although resonance can be found in all cases, no bound state can generated in this sector.

Interestingly, the scenario observed for $\Sigma(2455)$ recurs for $\Sigma(2620)$. Instead, $\Sigma(2620)$ falls within the resonance mass range corresponding to both the $\Sigma \phi$ and $\Xi K^\ast$ currents. Although the specific numerical results of these two currents differ, their quark configurations are both $qqss\bar{s}$ from the perspective of quark content.

We also know little about $\Sigma(2620)$, and it remains a one-star resonance in the RPP(PDG). However, our calculation results suggest that $\Sigma(2620)$ may contain a pentaquark component with the $qqss\bar{s}$ configuration, and mixing may exist between the color-singlet cluster combinations of the $\Sigma \phi$ and $\Xi K^\ast$. The possible spin parity assignment of $\Sigma(2620)$ should be $J^P=1/2^-$ suggested by our result.

What is more, the resonance generated by $N\phi$, $\Lambda \phi$, and $\Lambda K^\ast$ located near the two-body threshold of $N\phi(1680)$, $\Lambda(1600)\phi$, and $\Lambda(1810)K^*$, respectively, implies that these states may interpret as such molecules.

\begin{table*}[ht]
\renewcommand{\arraystretch}{1.5}
\caption{$J^P=1/2^+$ states generated from a baryon octet and a vector meson. The notation is the same as its in Table~\ref{tab_result_Pminus}.}
\centering
\resizebox{\textwidth}{!}{
\begin{tabular}{c|c|c|c|c|c|c}
Constituent hadrons  & $E_{th}$ (GeV) & $M_B^2$ (GeV$^2$) & $\sqrt{s_0}$ (GeV) & $M_H$ (GeV)& Molecule & Candidate\\\hline
$N_-\phi$ & $2.560$ & $2.0$--$2.7$ & $3.3\pm 0.1$ & $3.00\pm 0.09$ & $N(1895)\phi$ & $-$\\
$\Lambda_-\phi$ & $2.570$ & $\cdots$ & $\cdots$ & No state & $-$ & $-$\\
$\Sigma_-\phi$ & $2.650$ & $2.0$--$2.7$ & $3.3\pm0.1$ & $3.01\pm0.10$ & $\Sigma(1900)\phi$ & $-$\\
$\Xi_-\phi$ & $2.650$ & $\cdots$ & $\cdots$ & No state & $-$ & $-$\\
$\Lambda_- K^\ast$ & $2.440$ & $\cdots$ & $\cdots$ &  No state & $-$ & $-$\\
$\Sigma_- K^\ast$ & $2.520$ & $1.9$--$2.5$ & $3.1\pm0.1$ & $2.82\pm0.09$ & $\Sigma K^\ast(1680)$ & $-$\\
$\Xi_- K^\ast$ & $2.520$ & $\cdots$ & $\cdots$ & No state & $-$ & $-$
\end{tabular}
}
\label{tab_result_Vplus}
\end{table*}

Similarly, we can continue to discuss the situation when we project onto the positive-parity states. We still apply the masses listed in Table \ref{tab_minus_mass} to determine our two-body thresholds and constrain our threshold parameter $\sqrt{s_0}$ ranges. The calculated results for the corresponding $1/2^+$ states generated from a baryon octet and a vector meson are listed in Table \ref{tab_result_Vplus}.

The same situation occurs once again. For the positive-parity sector, we either fail to obtain a corresponding resonance, or the mass of the resonance we obtain is extremely large. Besides the $\Sigma_-K^*$ current obtain a resonance near $2.8$ GeV, all other channels where resonances can be obtained, we find only states with central masses above $3.0$ GeV.

\subsection{$1/2^+$ and $1^-$ to $J = 3/2$ system}

\begin{table*}[ht]
\renewcommand{\arraystretch}{1.5}
\caption{$J^P=3/2^-$ states generated from a baryon octet and a vector meson. The notation is the same as its in Table~\ref{tab_result_Pminus}.}
\centering
\resizebox{\textwidth}{!}{
\begin{tabular}{c|c|c|c|c|c|c}
Constituent hadrons  & $E_{th}$ (GeV) & $M_B^2$ (GeV$^2$) & $\sqrt{s_0}$ (GeV) & $M_H$ (GeV)& Molecule & Candidate\\\hline
$N\phi$ & $1.958$ & $1.6$--$2.2$ & $2.6\pm0.1$ & $2.37\pm0.07$ &  $-$ & $-$\\
$\Lambda\phi$ & $2.135$ & $2.2$--$2.8$ & $3.0\pm0.1$ & $2.73\pm0.08$ & $\Lambda\phi(1680)$ & $-$\\
$\Sigma\phi$ & $2.209$ & $1.7$--$2.1$ & $2.6\pm0.1$ & $2.31\pm0.09$ & $-$ & $\Sigma(2250)$ \\
$\Xi\phi$ & $2.334$ & $2.3$--$2.8$ & $3.0\pm0.1$ & $2.66\pm0.11$ & $-$ & $-$\\
$\Lambda K^\ast$ & $2.007$ & $1.6$--$2.2$ & $2.5\pm0.1$ & $2.23\pm0.07$ & $-$ & $-$\\
$\Sigma K^\ast$ & $2.081$ & $2.2$--$2.8$ & $3.1\pm0.1$ & $2.88\pm0.08$ & $-$ & $-$\\
$\Xi K^\ast$ & $2.206$ & $1.7$--$2.1$ & $2.6\pm0.1$ & $2.30\pm0.10$ & $-$ & $\Sigma(2250)$
\end{tabular}}
\label{tab_result_V3minus}
\end{table*}

 As shown in Table~\ref{tab_result_V3minus}, all cases related to $J^P=3/2^-$ states generated from a baryon octet and a vector meson are listed.
 In the this sector, the scenario observed for $\Sigma(2455)$ and $\Sigma(2620)$ recurs for $\Sigma(2250)$ again. $\Sigma(2250)$ also falls within the resonance mass range corresponding to both the $\Sigma \phi$ and $\Xi K^\ast$ currents. Although the specific numerical results of these two currents differ, their quark configurations are both $qqss\bar{s}$ from the perspective of quark content. 
With a two-star resonance in the RPP(PDG) $\Sigma(2250)$, our calculation results suggest that it should contain a pentaquark component with the $qqss\bar{s}$ configuration, and mixing may exist between the color-singlet cluster combinations of the $\Sigma \phi$ and $\Xi K^\ast$, while the possible spin parity assignment of $\Sigma(2250)$ should be $J^P=3/2^-$.
Moreover, considering the uncertainties in the computed masses, the \(\Xi K^*\) system with \(J^P = 3/2^-\) emerges as the only remaining candidate that may potentially form a bound-state solution within our calculation. A very recent result from BESIII claimed a state found in the $\bar{p} K^{+}$invariant mass of $\psi(3686) \rightarrow$ $\bar{p} K^{+} \Sigma^0$ channel, named $\Sigma(2330)$, with spin-parity $J^P=$ $3 / 2^{-}$~\cite{BESIII:2026glf}, which meets our predictions from the $\Xi K^*$ and $\Sigma \phi$ currents; further cross-check should be performed.

It is worth to be metioned that the constituent quark model predicts that $\Sigma K^\ast$ and $p\phi$ with $J^P=3/2^-$ can form bound states~\cite{Huang:2018ehi, Liu:2018nse}. Meanwhile, using the one-boson-exchange potential, bound states arising from $\Sigma K^\ast$ interactions are investigated via a quasipotential Bethe-Salpeter equation approach~\cite{He:2017aps}. Unfortunately, in our calculations of QCD sum rule performed with the currents we constructed, we did not obtain corresponding bound states in $\Sigma K^\ast$ and $N\phi$ channels.

\begin{table*}[ht]
\renewcommand{\arraystretch}{1.5}
\caption{$J^P=3/2^+$ states generated from a baryon octet and a vector meson. The notation is the same as its in Table~\ref{tab_result_Pminus}.}
\centering
\resizebox{\textwidth}{!}{
\begin{tabular}{c|c|c|c|c|c|c}
Constituent hadrons  & $E_{th}$ (GeV) & $M_B^2$ (GeV$^2$) & $\sqrt{s_0}$ (GeV) & $M_H$ (GeV)& Molecule & Candidate\\\hline
$N_-\phi$ & $2.560$ & $\cdots$ & $\cdots$ & No state & $-$ & $-$\\
$\Lambda_-\phi$ & $2.570$ & $2.0$--$2.7$ & $2.6\pm0.1$ & $2.08\pm0.08$ & $-$ & $\Lambda(2070)$\\
$\Sigma_-\phi$ & $2.650$ & $\cdots$ & $\cdots$ & No state & $-$ & $-$\\
$\Xi_-\phi$ & $2.650$ & $2.4$--$3.4$ & $2.8\pm0.1$ & $2.24\pm0.09$ & $-$ & $\Xi(2250)$\\
$\Lambda_- K^\ast$ & $2.440$ & $\cdots$ & $\cdots$ & No state & $-$ & $-$\\
$\Sigma_- K^\ast$ & $2.520$ & $2.0$--$3.0$ & $2.5\pm0.1$ & $2.00\pm0.08$ & $-$ & $N(2040)$\\
$\Xi_- K^\ast$ & $2.520$ & $\cdots$ & $\cdots$ & No state & $-$ & $-$
\end{tabular}
}
\label{tab_result_V3plus}
\end{table*}

When it comes to the positive-parity sector, the results of $J^P=3/2^+$ states generated from a baryon octet and a vector meson are listed in Table~\ref{tab_result_V3plus}. Two extremes emerge this time. We either find a deeply bound state very easily, or we fail to find a corresponding resonance at all. For $\Lambda_-\phi$, $\Xi_-\phi$, and $\Sigma_-K^\ast$ currents, bound states with a very large binding energy of hundreds of MeV are obtained, corresponding to $\Lambda(2070)3/2^+$, $\Xi(2250)$, and $N(2040)3/2+$ states listed in RPP(PDG). 
The results here lead to a very interesting interpretation, that is while these states are composed of two color-singlet clusters, for some reason, these two color-singlet clusters can bind unexpectedly tightly.
In addition, our calculation suggest the spin parity assignment of $\Xi(2250)$ should be $3/2^+$.

\subsection{Mass splitting}

\begin{table*}[ht]
\renewcommand{\arraystretch}{1.5}
\caption{$J=1/2$ spectrum generated from a baryon octet and a pseudoscalar meson via different Lorentz structure $\Pi_1$ and $\Pi_2$. Where constituent hadrons refer to the specific quark configurations, $E_{th}$ is the two-particle threshold of constituent hadrons, $M^2_B$ is the energy range of the Borel window, $\sqrt{s_0}$ is the threshold parameter, and $M_H$ is the calculated mass.}
\centering
\resizebox{\textwidth}{!}{
\begin{tabular}{c|c|c|c|c|c|c|}
 Lorentz structures& \multicolumn{3}{c|}{$\Pi_1$} & \multicolumn{3}{c|}{$\Pi_2$} \\\hline
 Constituent hadrons & $M_B^2$ (GeV$^2$) & $\sqrt{s_0}$ (GeV) & $M_H$ (GeV)& $M_B^2$ (GeV$^2$) & $\sqrt{s_0}$ (GeV) & $M_H$ (GeV)\\\hline
$N\eta_s$ & $2.1$--$2.6$ & $2.8\pm 0.1$ & $2.44\pm 0.09 $ & $2.0$--$2.6$ & $2.6\pm 0.1$ & $2.36\pm 0.08 $\\
$\Lambda\eta_s$  & $2.0$--$2.5$ & $2.7\pm 0.1$ & $2.48\pm 0.08$ & $2.0$--$2.6$ & $2.7\pm 0.1$ & $2.38\pm 0.09$\\
$\Sigma\eta_s$ & $2.3$--$2.9$ & $3.0\pm 0.1$ & $2.55\pm 0.11$ & $2.1$--$2.7$ & $2.7\pm 0.1$ & $2.39\pm 0.09$ \\
$\Xi\eta_s$ & $2.0$--$2.6$ & $2.8\pm 0.1$ & $2.54\pm 0.08$ & $1.8$--$2.6$ & $2.6\pm 0.1$ & $2.30\pm 0.07$\\
$\Lambda K$ & $2.2$--$2.8$ & $2.9\pm 0.1$ & $2.62\pm 0.08$ & $2.2$--$2.8$ & $2.8\pm 0.1$ & $2.45\pm 0.09$\\
$\Sigma K$  & $2.0$--$2.7$ & $2.8\pm 0.1$ & $2.45\pm 0.09$ & $2.2$--$2.8$ & $2.7\pm 0.1$ & $2.39\pm 0.08$\\
$\Xi K$& $2.0$--$2.7$ & $2.8\pm 0.1$ & $2.54\pm 0.08$ & $2.0$--$2.6$ & $2.7\pm 0.1$ & $2.38\pm 0.08$
\end{tabular}
}
\label{tab_result_P}
\end{table*}

\begin{table*}[ht]
\renewcommand{\arraystretch}{1.5}
\caption{$J=1/2$ spectrum generated from a baryon octet and a vector meson via different Lorentz structures $\Pi_1$ and $\Pi_2$. The notation is the same as in Table~\ref{tab_result_P}.}
\centering
\resizebox{\textwidth}{!}{
\begin{tabular}{c|c|c|c|c|c|c|}
 Lorentz structures & \multicolumn{3}{c|}{$\Pi_1$} & \multicolumn{3}{c|}{$\Pi_2$} \\\hline
 Constituent hadrons & $M_B^2$ (GeV$^2$) & $\sqrt{s_0}$ (GeV) & $M_H$ (GeV)& $M_B^2$ (GeV$^2$) & $\sqrt{s_0}$ (GeV) & $M_H$ (GeV)\\\hline
$N\phi$ & $2.2$--$2.8$ & $3.0\pm0.1$ & $2.75\pm0.08$ & $2.2$--$2.7$ & $2.7\pm0.1$ & $2.46\pm0.08$ \\
$\Lambda\phi$ & $2.3$--$2.9$ & $3.1\pm0.1$ & $2.91\pm0.08$ & $2.1$--$2.7$ & $2.7\pm0.1$ & $2.45\pm0.08$ \\
$\Sigma\phi$ & $2.6$--$3.0$ & $3.1\pm0.1$ & $2.74\pm0.09$ & $2.1$--$2.7$ & $2.6\pm0.1$ & $2.39\pm0.07$ \\
$\Xi\phi$ & $2.4$--$2.9$ & $3.1\pm0.1$ & $2.89\pm0.08$ & $2.0$--$2.6$ & $2.6\pm0.1$ & $2.38\pm0.07$ \\
$\Lambda K^\ast$ & $2.3$--$2.9$ & $3.1\pm0.1$ & $2.88\pm0.08$ & $2.3$--$2.9$ & $2.9\pm0.1$ & $2.57\pm0.09$ \\
$\Sigma K^\ast$ & $2.2$--$2.8$ & $3.0\pm0.1$ & $2.75\pm0.08$ & $2.3$--$2.8$ & $2.8\pm0.1$ & $2.53\pm0.08$\\
$\Xi K^\ast$ & $2.3$--$2.9$ & $3.1\pm0.1$ & $2.87\pm0.08$ & $2.1$--$2.6$ & $2.7\pm0.1$ & $2.46\pm0.08$
\end{tabular}
}
\label{tab_result_V1}
\end{table*}

\begin{table*}[ht]
\renewcommand{\arraystretch}{1.5}
\caption{$J=3/2$ spectrum generated from a baryon octet and a vector meson via different Lorentz structure $\Pi_1$ and $\Pi_2$. The notation is the same as its in Table~\ref{tab_result_P}.}
\centering
\resizebox{\textwidth}{!}{
\begin{tabular}{c|c|c|c|c|c|c|}
Lorentz structures & \multicolumn{3}{c|}{$\Pi_1$} & \multicolumn{3}{c|}{$\Pi_2$} \\\hline
Constituent hadrons & $M_B^2$ (GeV$^2$) & $\sqrt{s_0}$ (GeV) & $M_H$ (GeV)& $M_B^2$ (GeV$^2$) & $\sqrt{s_0}$ (GeV) & $M_H$ (GeV)\\\hline
$N\phi$ & $2.5$--$3.0$ & $3.1\pm0.1$ & $2.77\pm0.09$ & $2.0$--$2.6$ & $2.6\pm0.1$ & $2.23\pm0.08$ \\
$\Lambda\phi$ & $6.0$--$7.0$ & $2.6\pm0.1$ & $2.05\pm0.09$ & $2.0$--$2.7$ & $2.5\pm0.1$ & $2.09\pm0.09$ \\
$\Sigma\phi$ & $2.1$--$2.8$ & $2.9\pm0.1$ & $2.63\pm0.08$ & $2.0$--$2.6$ & $2.6\pm0.1$ & $2.19\pm0.08$\\
$\Xi\phi$ & $7.5$--$8.5$ & $2.8\pm0.1$ & $2.23\pm0.09$ & $2.2$--$2.9$ & $2.7\pm0.1$ & $2.19\pm0.09$\\
$\Lambda K^\ast$ & $2.3$--$2.9$ & $3.0\pm0.1$ & $2.75\pm0.08$ & $1.5$--$2.3$ & $2.3\pm0.1$ & $1.98\pm0.07$ \\
$\Sigma K^\ast$ & $6.0$--$7.0$ & $2.5\pm0.1$ & $1.99\pm0.09$ & $2.0$--$2.6$ & $2.5\pm0.1$ & $2.04\pm0.08$ \\
$\Xi K^\ast$ & $2.4$--$3.0$ & $3.2\pm0.1$ & $2.83\pm0.09$ & $1.7$--$2.6$ & $2.6\pm0.1$ & $2.15\pm0.09$
\end{tabular}
}
\label{tab_result_V3}
\end{table*}

Recall that the $\Pi_1$ and $\Pi_2$ corresponding to the chiral-even and chiral-odd structures in the correlation functions, $\Pi_1$ contains positive and negative states with the same sign, while $\Pi_2$ contains them with different signs. To figure out the possible relationship, we separately calculated the so-called hadronic mass corresponding to the different Lorentz structure $\Pi_1$ and $\Pi_2$. The extracted $J=1/2$ spectrum generated from a baryon octet and a pseudoscalar meson, and the $J=1/2$ spectrum generated from a baryon octet and a vector meson, while the $J=3/2$ spectrum generated from a baryon octet and a vector meson via different Lorentz structure $\Pi_1$ and $\Pi_2$ are listed in Tables~\ref{tab_result_P}, \ref{tab_result_V1}, and \ref{tab_result_V3}, respectively.

\begin{table}[htbp]
\renewcommand{\arraystretch}{1.5}
\centering
\caption{Comparison of the calculated hadron masses $M_H$ (in GeV) for the $J^P = 1/2$ sector generated from a baryon octet and a pseudoscalar meson, derived from different Lorentz structures $\Pi_1$, $\Pi_2$, $\Pi_-$, and $\Pi_+$. The table includes the mass differences $\Pi_+ - \Pi_-$ and $\Pi_1 - \Pi_2$.}
\label{tab:mass_comparison}
\begin{tabular}{c|c|c|c|c|c|c}
$M_H$ (GeV) & ~~~~$\Pi_1$~~~~ & ~~~~$\Pi_2$~~~~ & $\Pi_1 - \Pi_2$ & ~~~~$ \Pi_+$~~~~ & ~~~~$ \Pi_-$~~~~ & $\Pi_+ - \Pi_-$ \\ \hline
$N\eta_s$ & $2.44$ & $2.36$ & $0.08$ & $2.78$ & $2.42$ & $0.36$ \\
$\Lambda\eta_s$ & $2.48$ & $2.38$ & $0.10$ & $\cdots$ & $2.48$ & $\cdots$ \\
$\Sigma\eta_s$ & $2.55$ & $2.39$ & $0.14$ & $2.79$ & $2.38$ & $0.41$ \\ 
$\Xi\eta_s$ & $2.54$ & $2.30$ & $0.24$ & $\cdots$ & $2.41$ & $\cdots$ \\ 
$\Lambda K$ & $2.62$ & $2.45$ & $0.17$ & $\cdots$ & $2.49$ & $\cdots$ \\ 
$\Sigma K$ & $2.45$ & $2.39$ & $0.06$ & $2.76$ & $2.42$ & $0.34$ \\
$\Xi K$ & $2.54$ & $2.38$ & $0.16$ & $\cdots$ & $2.48$ & $\cdots$ \\ 
\end{tabular}
\end{table}
\begin{table}[htbp]
\renewcommand{\arraystretch}{1.5}
\centering
\caption{Comparison of the calculated hadron masses $M_H$ (in GeV) for the $J^P = 1/2$ sector generated from a baryon octet and a vector meson. Notation is the same as Table~\ref{tab:mass_comparison}}
\label{tab:mass_comparison_V1minus}
\begin{tabular}{c|c|c|c|c|c|c}
$M_H$ (GeV) & ~~~~$\Pi_1$~~~~ & ~~~~$\Pi_2$~~~~ & $\Pi_1 - \Pi_2$ & ~~~~$ \Pi_+$~~~~ & ~~~~$ \Pi_-$~~~~ & $\Pi_+ - \Pi_-$ \\ \hline
$N\phi$ & $2.75$ & $2.46$ & $0.29$ & $3.00$ & $2.67$ & $0.33$ \\
$\Lambda\phi$ & $2.91$ & $2.45$ & $0.46$ & $\cdots$ & $2.69$ & $\cdots$ \\
$\Sigma\phi$ & $2.74$ & $2.39$ & $0.35$ & $3.01$ & $2.56$ & $0.45$ \\
$\Xi\phi$ & $2.89$ & $2.38$ & $0.51$ & $\cdots$ & $2.62$ & $\cdots$ \\
$\Lambda K^\ast$ & $2.88$ & $2.57$ & $0.31$ & $\cdots$ & $2.70$ & $\cdots$ \\
$\Sigma K^\ast$ & $2.75$ & $2.53$ & $0.22$ & $2.82$ & $2.64$ & $0.18$ \\
$\Xi K^\ast$ & $2.87$ & $2.46$ & $0.41$ & $\cdots$ & $2.59$ & $\cdots$ \\
\end{tabular}
\end{table}

\begin{table}[htbp]
\renewcommand{\arraystretch}{1.5}
\centering
\caption{Comparison of the calculated hadron masses $M_H$ (in GeV) for the $J^P = 3/2$ sector generated from a baryon octet and a vector meson, derived from different Lorentz structures $\Pi_1$, $\Pi_2$, $\Pi_-$, and $\Pi_+$. The table includes the mass differences $\Pi_+ - \Pi_-$ and $\Pi_1 - \Pi_2$.}
\label{tab:mass_comparison_V3minus}
\begin{tabular}{c|c|c|c|c|c|c}
$M_H$ (GeV) & ~~~~$\Pi_1$~~~~ & ~~~~$\Pi_2$~~~~ & $\Pi_1 - \Pi_2$ & ~~~~$ \Pi_+$~~~~ & ~~~~$ \Pi_-$~~~~ & $\Pi_+ - \Pi_-$ \\ \hline
$N\phi$ & $2.77$ & $2.23$ & $0.54$ & $\cdots$ & $2.37$ & $\cdots$ \\
$\Lambda\phi$ & $2.05$ & $2.09$ & $-0.04$ & $2.08$ & $2.73$ & $-0.65$ \\
$\Sigma\phi$ & $2.63$ & $2.19$ & $0.44$ & $\cdots$ & $2.19$ & $\cdots$ \\
$\Xi\phi$ & $2.23$ & $2.19$ & $0.04$ & $2.24$ & $2.59$ & $-0.35$ \\
$\Lambda K^\ast$ & $2.75$ & $1.98$ & $0.77$ & $\cdots$ & $2.23$ & $\cdots$ \\
$\Sigma K^\ast$ & $1.99$ & $2.04$ & $-0.05$ & $2.00$ & $2.88$ & $-0.88$ \\
$\Xi K^\ast$ & $2.83$ & $2.15$ & $0.68$ & $\cdots$ & $2.20$ & $\cdots$ \\
\end{tabular}
\end{table}

To further directly shows the mass splitting, we listed the mass differences of above three cases in Tables~\ref{tab:mass_comparison}, \ref{tab:mass_comparison_V1minus}, and \ref{tab:mass_comparison_V3minus}. We found that for the $1/2$ and $3/2$ cases, the directions of their mass splittings are exactly opposite. Furthermore, for the cases where a resonance can be found for positive-parity, the difference between the $M_H$ values calculated by $\Pi_1$ and $\Pi_2$ is smaller than in those cases where no positive-parity resonance is found. From a theoretical perspective, the mass splitting itself is caused by chiral-odd terms, which are proportional to the quark condensate $\langle \bar{q}q \rangle$ and mixed condensate $\langle \bar{q}g \sigma \cdot G q \rangle$ terms~\cite{Oka:1996zz, Jido:1996ia}, which mainly correspond to the Lorentz structures of $\Pi_2$. When the difference between the $M_H$ values calculated by $\Pi_1$ and $\Pi_2$ is relatively large, which often corresponds to the mass of the $\Pi_2$ term being significantly lower than the mass of $\Pi_1$ term, it implies that the $\Pi_2$ term dominates the calculation. This results in an excessively large splitting between the positive- and negative-parity states, making it difficult for us to find the state corresponding to positive-parity within an appropriate $\sqrt{s_0}$ range.

\section{Summary and Conclusions} \label{sec:summary}

In this work, we have systematically studied hidden- and open-strange pentaquark configurations in the hadronic molecular picture by means of QCD sum rules. A total of 21 interpolating currents with $J=1/2$ and $3/2$ were constructed from the baryon octet combined with an $s\bar{s}$ pair. In the numerical analysis, we included OPE contributions up to dimension 11, considered both parity-projected spectra and the separate $\Pi_1$ and $\Pi_2$ structures, and incorporated the uncertainty associated with the factorization approximation of higher-dimensional condensates.

Our results show that no bound-state solution is found in the $J^P=1/2^-$ sector, while several channels generate resonance masses close to nearby thresholds or to poorly established strange baryon candidates. In the $J^P=3/2^-$ sector, resonance solutions are obtained, but still none support a bound-state interpretation.

In positive-parity sectors, including both $J^P = 1/2^+$ and $3/2^+$, it is more difficult to identify the corresponding states of the current we contracted. A large mass splitting was found between the positive- and negative-parity states raised from the same current.

Few spin-parity quantum number to be suggest by our calculation, including  $\Sigma(2250)$, $\Sigma(2455)$, $\Sigma(2620)$, $\Xi(2250)$, and $\Xi(2370)$, to be $J^P = 3/2^-$, $1/2^-$, $1/2^-$, $3/2^+$, and $1/2^-$, respectively. Meanwhile, these states are very favorable to the pentaquark configuration, especially for the $\Sigma$ states above. For all three $\Sigma$ states, it is found that they may contain a pentaquark component with the $qqss\bar{s}$ configuration, and mixing may exist between the color-singlet cluster combinations of $\Sigma+\eta_s/\phi$ and $\Xi + K/K^\ast$ currents.
The mixing between the currents will require further investigation in future work. This also seems to suggest that it may be possible to search for the corresponding pole structures within such coupled-channel frameworks.

Further studies with a more complete current basis may help clarify the nature of these states and their possible multiplet relations in the strange sector.
Alternative constructions may offer new possibilities for identifying such states. For instance, in future work, we will extend the construction of molecular currents from the baryon octet to the baryon decuplet. In this context, we will not only cover the range of spins from $1/2$ to $5/2$, but also consider the mixing between channels for specific quantum numbers. Future investigations into a broader range of current structures may provide further insights into the open questions addressed in this work, while also improving the interpretation of the light-flavor hadron spectrum from the perspective of QCD sum rules.

{\bf ACKNOWLEDGMENTS}

The authors appreciate the meaningful discussion with Professor Bing-Dong Wan and Professor Feng-Kun Guo. This work was supported by Tsinghua University, by the National Natural Science Foundation of China(NSFC) under Grants No. 12547114 and No. 12325503, and by the High-performance Computing Platform of Peking University.

{\bf DATA AVAILABILITY}

There are no publicly available research data or software supporting this manuscript. Requests for further information or data should be sent to the authors.

\appendix

\section{Appendix: Figures of the pole contribution, the OPE convergence and the mass}\label{app1}

\begin{figure*}[ht]
\centering
\includegraphics[width=0.8\textwidth]{R-Pi1.pdf}
\caption{The ratio with respect to the Borel parameter $M_B^2$ with different threshold parameter $\sqrt{s_0}$ for $\Pi_1$.}
\label{fig:R-Pi1}
\end{figure*}

\begin{figure*}[ht]
\centering
\includegraphics[width=0.8\textwidth]{M-Pi1.pdf}
\caption{The mass with respect to the Borel parameter $M_B^2$ with different threshold parameter $\sqrt{s_0}$ for $\Pi_1$.}
\label{fig:M-Pi1}
\end{figure*}

\begin{figure*}[ht]
\centering
\includegraphics[width=0.8\textwidth]{R-Pi2.pdf}
\caption{The ratio with respect to the Borel parameter $M_B^2$ with different threshold parameter $\sqrt{s_0}$ for $\Pi_2$.}
\label{fig:R-Pi2}
\end{figure*}

\begin{figure*}[ht]
\centering
\includegraphics[width=0.8\textwidth]{M-Pi2.pdf}
\caption{The mass with respect to the Borel parameter $M_B^2$ with different threshold parameter $\sqrt{s_0}$ for $\Pi_2$.}
\label{fig:M-Pi2}
\end{figure*}

\begin{figure*}[ht]
\centering
\includegraphics[width=0.8\textwidth]{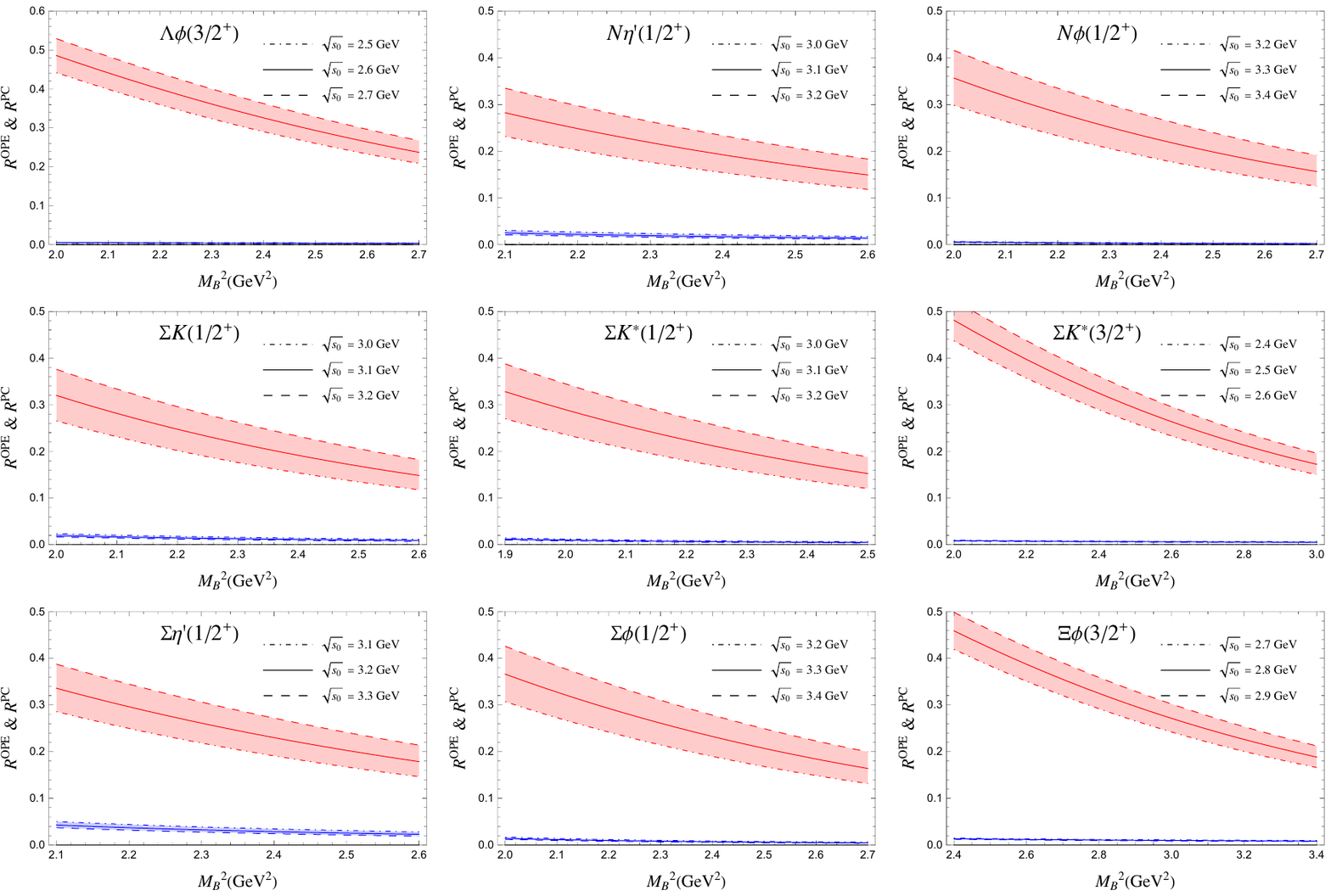}
\caption{The ratio with respect to the Borel parameter $M_B^2$ with different threshold parameter $\sqrt{s_0}$ for $\Pi_{+}$.}
\label{fig:R-Piplus}
\end{figure*}

\begin{figure*}[ht]
\centering
\includegraphics[width=0.8\textwidth]{R-Piminus.pdf}
\caption{The ratio with respect to the Borel parameter $M_B^2$ with different threshold parameter $\sqrt{s_0}$ for $\Pi_{-}$.}
\label{fig:R-Piminus}
\end{figure*}

\begin{figure*}[ht]
\centering
\includegraphics[width=0.8\textwidth]{M-Piminus.pdf}
\caption{The mass with respect to the Borel parameter $M_B^2$ with different threshold parameter $\sqrt{s_0}$ for $\Pi_{-}$.}
\label{fig:M-Piminus}
\end{figure*}

\end{document}